\documentclass{article}

\usepackage{graphicx, amssymb}

	\renewcommand {\i}[1] {_\mathrm{#1}}
	\newcommand {\e}[1] {^\mathrm{#1}}	
	\newcommand {\tens}[1] {\underline{#1}}
	 \newcommand {\TENS}[1] {\mathbb{#1}}
	\renewcommand {\d} {\mathrm{d}}
	\newcommand {\trace} {\mathrm{tr}}
		\newcommand {\ID} {\tens{I}}
		\newcommand {\SIG} {\tens{\sigma}}
		\newcommand {\SIGD} {\tens{\sigma}\e{d}}
		\newcommand {\SIGH} {\tens{\sigma}\e{h}}
		\newcommand {\EPS} {\tens{\varepsilon}}
		\newcommand {\EPSD} {\tens{\varepsilon}\e{d}}
		\newcommand {\EPSH} {\tens{\varepsilon}\e{h}}
		\newcommand {\eh} {\varepsilon \e h}
		\newcommand {\ed} {\varepsilon \e d}
		\newcommand {\sh} {\sigma \e h}
		\newcommand {\sd} {\sigma \e d}
		\newcommand {\si} {\sigma \i {I}}
		\newcommand {\sii} {\sigma \i {II}}
		\newcommand {\siii} {\sigma \i {III}}
		\newcommand {\nut} {\tilde{\nu}}
		\newcommand {\nust} {\hat{\nu}}
		
		\newcommand {\so} {\sigma \i 0}
		\newcommand {\sy} {\sigma \i y}
		\newcommand {\sco} {\sigma \e c}
		\newcommand {\ste} {\sigma \e t}
		
\begin{document}
\title{A damage model with non-convex free energy for quasi-brittle materials}
\author{Marc Fran\c cois}
\maketitle


\begin{abstract}
A state coupling between the hydrostatic (volumetric) and deviatoric parts of the free energy is introduced in a damage mechanics model relevant for the quasi-brittle materials.
It is shown that it describes the large dilatancy of concrete under compression and the different localization angles and damage levels in tension and compression.
A simple isotropic description is used, although similar ideas can be extended to anisotropic damage.
The model is identified with respect to tensile and compression tests and validated on bi-compression and bi-tension.
Fully written in three dimensions under the framework of thermodynamics of irreversible processes, it allows further developments within a finite element code.
\end{abstract}


\section{Introduction}
At the ultimate state of damage, quasi-brittle materials can be regarded as granular materials whose behavior, due to grain interlock, is strongly dependent on the confinement. Classical yield criteria (Mohr-Coulomb, Drucker-Prager,...) use the confinement (the hydrostatic strain) as a reinforcement factor. But there is experimental evidence that elasticity of powder materials, thus their free energy, also depend on it \cite{ouglova_06}. That point is generally not taken into account in damage models. As a consequence, the large dilantancy observed at these states, for example on concretes \cite{jamet_84,kupfer_69}, is poorly described by classical damage models as apparent Poisson's ratio cannot exceed 0.5, the limit for linear elasticity, while experiments exhibit greater values.
Another specificity of concrete behavior is the very different crack angles observed in tension and compression. In most damage models, the Hadamart and Rice criterion \cite{rice_75,comi_95} leads to the same localization angle in both cases.
Furthermore, for a concrete specimen, even after a rupture in tension, some carrying capacity in compression remains: this implies a damage level at the onset of localization much lower in tension than in compression. 
The present model describes these effects that are generally missed by most of damage models.

This work, in continuity with \cite{le_05}, constitutes an attempt to use a non convex potential in the field of damage mechanics. For this reason, a simple isotropic damage law is considered. As in the Kelvin's approach of elasticity \cite{rychlewski_84}, plasticity theory, soil mechanics and some damage models \cite{ladeveze_93}, the isotropic and deviatoric decomposition is used.
The retained yield criterion \cite{francois_08} is smooth and convex.
The model is identified with respect to the well known uniaxial and multiaxial testings of \cite{kupfer_69}.
%
%
\section{Constitutive law}
%
%

\textbf{Helmholtz free energy.}
The damage level is described by the scalar variable $d$ that ranges from 0 for sound material to 1 for fully damaged material \cite{lemaitre_96}. The present model is an isotropic one: the hydrostatic and deviatoric partitions of the stress $\SIG=\SIGH+\SIGD$ and the strain $\EPS=\EPSH+\EPSD$ are used (details are given in Appendix \ref{tensoform}).
The state variables $(\EPSD, \EPSH, d)$ describe the material's state and, with respect to the generalized standard material framework \cite{halphen_75}, the associated thermodynamic forces are respectively $(\SIGD, \SIGH, Y)$ where $Y$ is  the energy release rate density. The proposed free energy $\Psi$ is:
\begin{equation}
2\rho \Psi(\EPSD, \EPSH, d) = 3K \EPSH:\EPSH
+ 2 \mu \left( 1- d (1 + 2 \varphi \eh ) \right) \EPSD:\EPSD,
\label{helmholtz}
\end{equation}
where $\eh=\trace (\EPS)/\sqrt{3}$ (Appendix \ref{tensoform}). The new constant introduced is $\varphi$, the other ones are the mass density $\rho$ and the bulk and shear moduli, respectively $K$ and $\mu$. The role of $\varphi$ will be detailed further but it can be already seen that it introduces a cubic term in the free energy and that setting $\varphi=0$ leads to a simple damage model in which only the deviatoric part is affected by damage. This choice has been made in order to get rid of the unilateral effect of damage on bulk modulus (that comes from crack opening and closure) because the use of positive parts of the strain tensor induces difficulties associated with non regular free energy \cite{badel_07}.
In Eq.~\ref{helmholtz} tensors $\EPSH$ and $\EPSD$ can be replaced by their  algebraic values $\eh$ and $\ed$ (Appendix \ref{tensoform}).
%
%

\textbf{Stress to strain relationship.}
The hydrostatic and deviatoric stresses, as thermodynamic forces, are obtained by differentiation of the free energy with respect to $\EPSH$ and $\EPSD$:
\begin{eqnarray}
\SIGH &=& 3K \EPSH - \frac{2}{\sqrt{3}} \mu \varphi d (\ed)^2 \ID\label{sigh},\\
\SIGD &=& 2 \mu \left( 1- d (1 + 2 \varphi \eh ) \right) \EPSD\label{sigd}.
\end{eqnarray}
Setting $d=0$ leads to recover the linear isotropic elasticity law in the Kelvin's decomposition form \cite{rychlewski_84}.
%
%
The deviatoric stress and strain remain collinear together (and collinear to the unitary tensor $\tens{D}$, see Appendix \ref{tensoform}). Once projected on the orthogonal tensor base $(\tens{I}, \tens{D})$, the previous expression becomes:
%
\begin{eqnarray}
\sh &=& 3 K \eh - 2 \mu \varphi d (\ed)^2\label{sigh1},\\
\sd &=& 2 \mu \left( 1- d - 2 \varphi d \eh \right) \ed\label{sigd1}.
\end{eqnarray}
%
Let us suppose an imposed deviatoric strain while the hydrostatic strain remains equal to zero (isochoric transformation) then a confining pressure ($\sh<0$) is necessary to keep the volume unchanged:
\begin{equation}
	\sh (\eh=0,\ed) = - 2 \mu \varphi d (\ed)^2 \label{eqnh1}.
\end{equation}
Let us suppose now that the deviatoric strain is joined with an hydrostatic stress imposed equal to zero then an induced dilatation $\eh>0$ arises:
\begin{equation}
	\eh (\sh=0,\ed) = \frac{2 \mu \varphi d (\ed)^2}{1-d} \label{eqnh2}.
\end{equation}
These two aspects are relevant to the dilatancy effect in concrete-like materials. The micro-mechanical point of view associated to these effects is the surmounting of concrete particles, inducing voids creation, that arises under irreversible shearing \cite{yazdani_88, francois_05}.
%
%

\textbf{Elastic tangent modulus.}
The tangent modulus is defined as $\TENS{H}=\d \SIG / \d \EPS$.
In case of elastic transformation, $d$ remains fixed, then $\TENS{H}$ reduces to $\TENS{H}^0=\partial \SIG / \partial \EPS$.
From the stress to strain relations (\ref{sigh}, \ref{sigd}), we have:
\begin{eqnarray}
\TENS{H}^0 &=& 
\frac{3K-2\tilde{\mu}}{3} \ID\otimes\ID + 2\tilde{\mu} \TENS{I}- \alpha \left( \ID \otimes \EPSD + \EPSD \otimes \ID \right)\label{H0},\\
2\tilde{\mu} &=& \frac{\partial \sd}{\partial \ed}
= 2 \mu (1- d) - 4 \mu \varphi d \eh \label{mutilde},\\
\alpha&=&\frac{4 \mu \varphi d}{\sqrt{3}} \label{alpha}.
\end{eqnarray}
In this expression $\TENS{I}$ is the fourth order identity whose expression in index form is $[\TENS{I}]\i{ijkl}=(\delta\i{ik}\delta\i{jl}+\delta\i{il}\delta\i{jk})/2$, the symbol $\otimes$ refers to the tensor product and (from Eq.~\ref{sigd1}) $\tilde{\mu}$ represents the apparent shear modulus. Then $\TENS{H}^0$ has the index symmetries of an elasticity tensor: $[\TENS{H}^0]\i{ijkl}=[\TENS{H}^0]\i{klij}=[\TENS{H}^0]\i{ijlk}$.
%

\textbf{Inverse stress to strain relationship.}
When $d\neq0$, (the case $d=0$ is straightforward), replacing $\eh$ in Eq.~\ref{sigd1} by its value (Eq.~\ref{sigh1}) gives, first, $\ed$ as the solutions of a third order equation and, second, the relation between $\eh$ and $\ed$:
\begin{eqnarray}
(\ed)^3 + p \ed + q &=& 0\nonumber,\\
p = \frac {2 \varphi d \sh-  3K(1-d)} {4\mu\varphi^2 d^2} &,&
q = \frac {3K \sd} {8\mu^2\varphi^2 d^2}\nonumber,\\
\eh &=& \frac{\sh + 2 \mu \varphi d \left( \ed \right) ^2}{3K} \label{inverse}.
\end{eqnarray}
The discriminant $\Delta = (p/3)^3 + (q/2)^2$ leads to analyse from one to three real solutions.  This non univocal inverse relationship would not be used in finite element calculation, but we shall have to consider these three possibilities for the semi-analytical resolutions in Sec.~\ref{sectenscomp} and \ref{secbi}.
%
%
\section{Free energy analysis}\label{freeeneranal}
\begin{figure}[htbp]
\begin{center}
\includegraphics[scale=0.4]{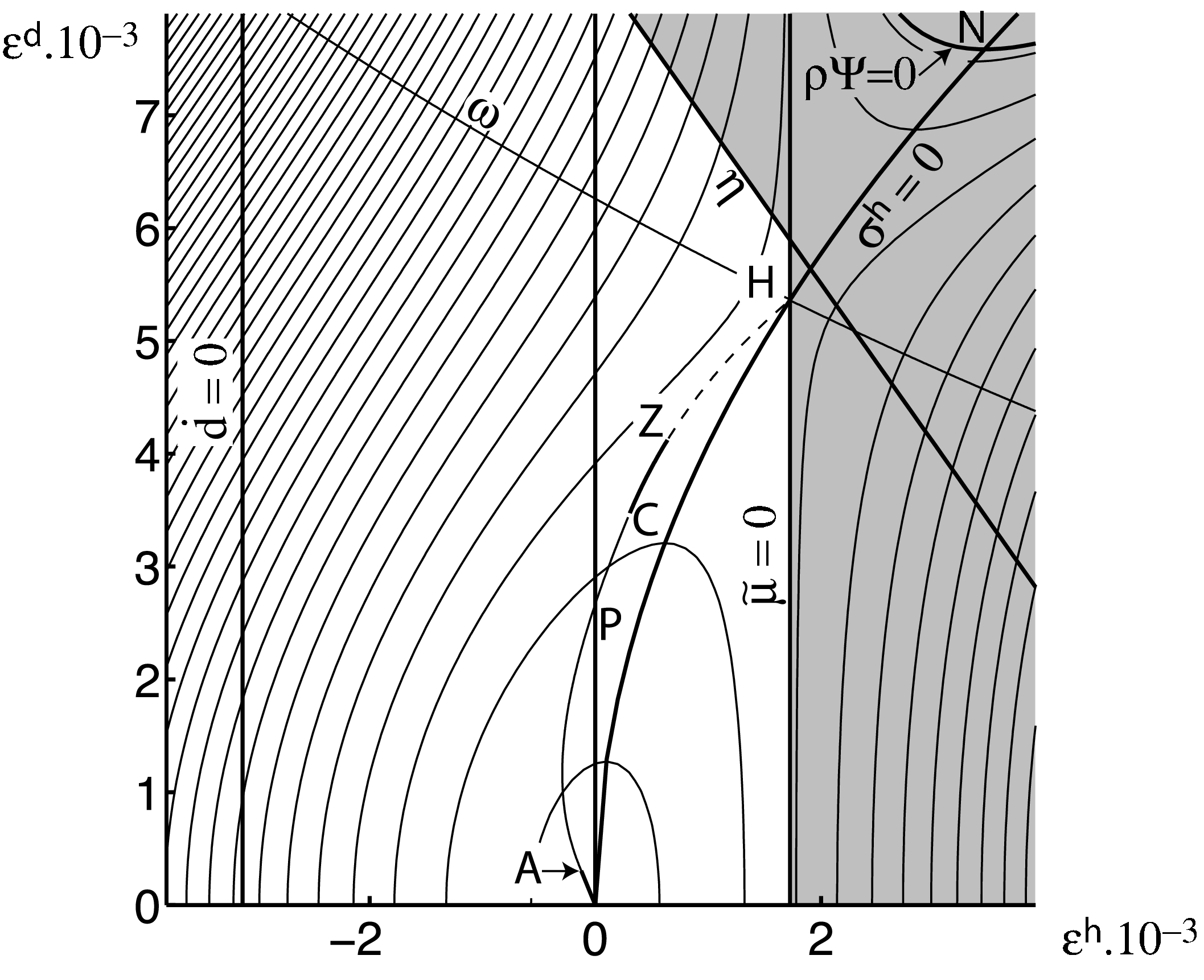}
\caption{Iso-values of the free energy in the Drucker-Prager plane for $d$=0.64}
\label{planDP}
\end{center}
\end{figure}

The proposed Helmholtz non convex free energy exhibits complex behavior that has to be carefully studied. Fig.~\ref{planDP} shows the contour plot of the free energy defined by Eq.~\ref{helmholtz}.
The transformation APCZ is the compression studied in Sec.~\ref{sectenscomp} and the damage value retained for the drawing is $d$(C), the maximum reached during this transformation.
%

\textbf{Positivity of the free energy.}
It signifies that the material cannot restore more energy than has been stored inside. In linear elasticity, it is associated with the positiveness of the Kelvin's moduli $(3K,2\mu)$ and infers classic bounds for the Poisson's ratio $(-1\leqslant\nu\leqslant0.5)$. In the general case, from Eq.~\ref{helmholtz} we have:
\begin{equation}
\rho \Psi<0 \Leftrightarrow \ed > \sqrt{-\frac{3K}{2\tilde{\mu}}} \eh.
\label{rhopsieqzero}
\end{equation}
This condition corresponds to the forbidden domain inside the curve $\rho \Psi=0$ in Fig.~\ref{planDP}. It only exists when $\tilde{\mu}<0$. It will be shown in Sec.~\ref{freeeneranal} that it cannot be reached during a transformation because localization occurs at least when $\tilde{\mu}=0$.
%

\textbf{Particular lines.}
Lines of interest are the loci of $\sd=0$ and $\sh=0$. From Eq.~\ref{sigd1} we have:
\begin{equation}
\eh = \frac{1-d}{2\varphi d} \Rightarrow \sd=0.
\label{mutltzero}
\end{equation}
From Eq.~\ref{mutilde}, this corresponds to $\tilde{\mu}=0$. On the right side of this line on Fig.~\ref{planDP}, $\tilde{\mu}<0$. 
The second solution for $\sd=0$ is more classically $\ed=0$. 
The locus of $\sh=0$ (Fig.~\ref{planDP}) is given by Eq.~\ref{sigh1}:
\begin{equation}
\ed=\sqrt{\frac{3K\eh}{2 \mu \varphi d}} \Rightarrow \sh=0.
\end{equation}
Between the deviatoric axis $\eh=0$ and this line, $\eh>0$ and $\sh<0$. This represents the dilatancy effect of the model: the deviatoric strain induces dilatancy even if the material is under (moderated) pressure. 
The saddle point H is defined as the intersection of the two lines $\sh=0$ and $\sd=0$, we have:
\begin{equation}
\eh(\textrm{H})= \frac{1-d}{2\varphi d},\quad
\ed(\textrm{H})= \frac{1}{2\varphi d}\sqrt{\frac{3K}{\mu}(1-d)},\quad
\rho\Psi(\textrm{H})= \frac{3K(1-d)^2}{8\varphi^2 d^2}.
\end{equation}
As $\SIG(H)=\tens{0}$, it corresponds to an unstable free stress state with $\EPS(H)\neq 0$. Finally, it can be easily shown that the point N, corresponding to the minimum deviatoric strain for $\rho\Psi=0$ is such as: $\eh(\textrm{N}) = 2\eh(\textrm{H})$, $\ed(\textrm{N}) = \sqrt{2}\ed(\textrm{H})$.
%

\textbf{Positivity of the energy release rate.}
The energy release rate $Y$ is associated with the damage $d$:
\begin{equation}
Y = - \frac{\d \rho \Psi}{\d d} = \mu (1+2\varphi \eh) (\ed)^2 \label{Y}.
\end{equation}
The thermal dissipation is $\dot{q}_i^d=Y \dot{d}$. As damage $d$ cannot physically decrease, its positiveness implies that if $Y$ is negative, the evolution of the damage must stop:
\begin{equation}
\eh < \frac{-1}{2\varphi} \Rightarrow \dot{d}=0.
\label{limitY}
\end{equation}
In the example of Sec.~\ref{sectenscomp}, $\varphi=160$ then this condition is $\eh <-3.12\,10^{-3}$. That region, denoted as $\dot{d}=0$ on Fig.~\ref{planDP} corresponds to high confinements out of most practical applications (far to be reached in the presented examples); it is strongly due to the initial choice of damage acting only on the hydrostatic part. One can remark that some experiments tend to confirm the existence of a damage locking at high confinements \cite{burlion_97}.
%

\textbf{Localization.}
The Hadamart and Rice criterion \cite{rice_75,comi_95} of localization authorizes the existence of a localization plane (a concentration of large strains) orthogonal to the unit vector $\vec{n}$ as soon as:
\begin{equation}
\det(\vec{n}.\TENS{H}.\vec{n})\leqslant 0\label{detnhn}.
\end{equation}
Although we still consider here elastic transformations, we assume this localization to correspond to the apparition of a macroscopic crack. Once appeared, the damage model ceases to apply as the body is split into pieces: it infers a restriction to the domain of definition of the model.
Among possible vectors $\vec{n}$, we consider $\vec{n}\i{I}$, an eigenvector of the strain deviator tensor, then $\EPSD.\vec{n}\i{I}=\varepsilon\e{d}\i{I}\vec{n}\i{I}$, where $\varepsilon\e{d}\i{I}$ is the corresponding eigenvalue of $\EPSD$ (a principal deviatoric strain). As $\TENS{H}=\TENS{H}^0$ in this case, Eq.~\ref{H0} give:
\begin{equation}
\vec{n}\i{I}.\TENS{H}^0.\vec{n}\i{I} = 
\left( \frac{3K+\tilde{\mu}}{3} -2\alpha\varepsilon\e{d}\i{I} \right) \vec{n}\i{I}\otimes\vec{n}\i{I} + \tilde{\mu}\ID.
\end{equation}
The determinant of this expression is:
\begin{equation}
\det(\vec{n}\i{I}.\TENS{H}^0.\vec{n}\i{I})=\tilde{\mu}^2\left( \frac{3K+4\tilde{\mu}}{3} -2\alpha\ed\i{I} \right).
\end{equation}
This shows that $\tilde{\mu}=0$ is a sufficient condition of localization.
The domain (grayed) on the right side of this line in Fig.~\ref{planDP} will not be concerned by the possible evolutions.
From Eq.~\ref{rhopsieqzero}, it always contains the domain $\rho\Psi\leqslant0$.
%
From Eq.~\ref{epsbound}, $\ed\i{I}\geqslant\ed/\sqrt{6}\geqslant0$, considering that $K$, $\tilde{\mu}$ and $\alpha$ are positive and using Eq.~\ref{mutilde} and \ref{alpha}, we obtain a second sufficient condition for localization:
%
\begin{equation}
	\ed\geqslant\frac{3K+4\mu(1-d)}{4\sqrt{2}\mu\varphi d}-\sqrt{2}\eh.
\end{equation}
The corresponding line is denoted as $\eta$ on the Fig.~\ref{planDP}. 
The localization occurs at least when a transformation reaches the lines $\eta$ or $\tilde{\mu}=0$; the domain above them (grayed) is not concerned by possible evolutions.
%

\textbf{Other remarks about stability.}
The current line $\omega$ passes by the saddle point H. From its definition $\dot{\eh} / \sh = \dot{\ed} / \sd$ comes:
\begin{equation}
2 \mu (1- d)  \ed \dot{\eh} - 4 \mu \varphi d \eh \ed \dot{\eh} = 3 K \eh \dot{\ed} - 2 \mu \varphi d (\ed)^2  \dot{\ed}.
\end{equation}
This expression has no simple analytic solution and $\omega$ has been drawn on Fig.~\ref{planDP} with a steepest slope algorithm. Below $\omega$, the stress tends to bring back the system to the stable original state 0; above $\omega$ it tends to make the system reach the domain above the line $\eta$ where the localization occurs. Then the domain between $\omega$ and $\eta$ can be considered as unstable, leading to an instantaneous evolution towards the lines $\eta$ and $\tilde{\mu}=0$ where the localization occurs.
%
%

%
%
%
\section{Dissipative behaviour}\label{dissbe}
%
%

\textbf{Yield surface.}
\begin{figure}[htbp]
\begin{center}
\includegraphics[scale=0.5]{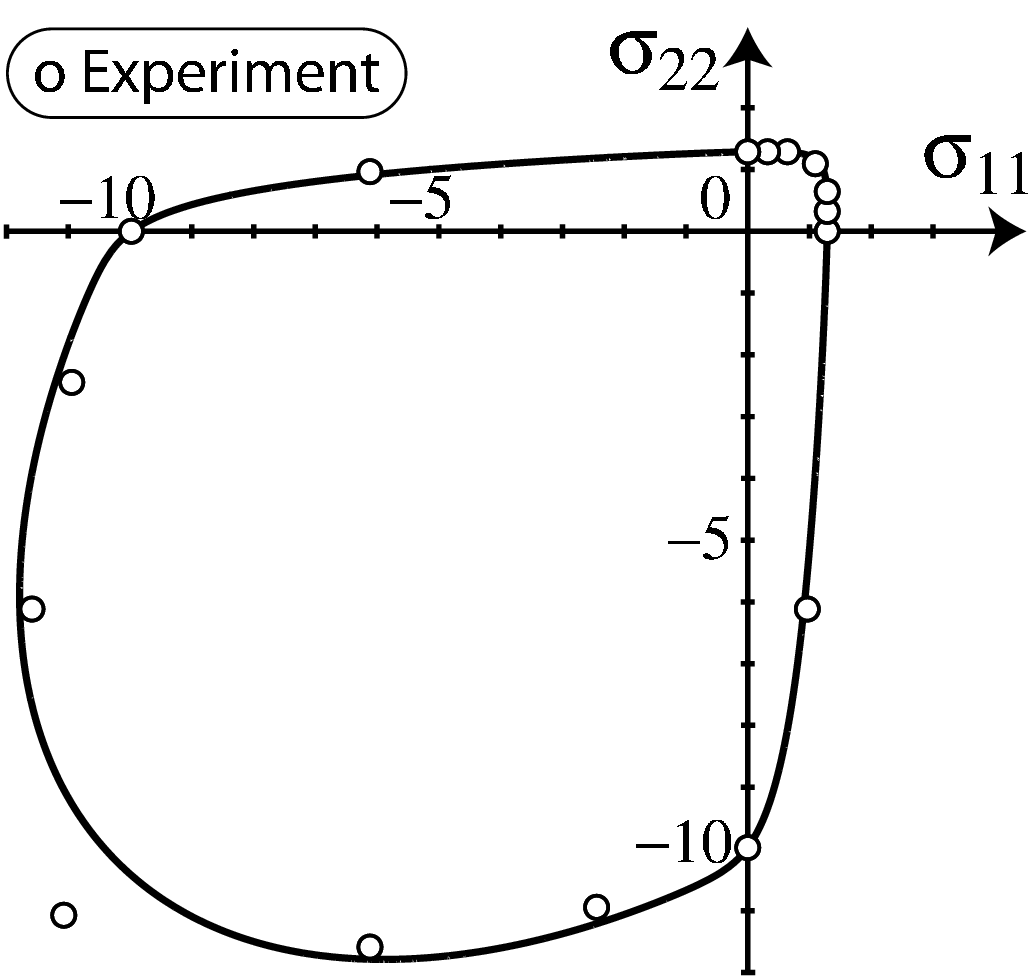}
\caption{Initial yield surface (MPa)}
\label{SSFrancois}
\end{center}
\end{figure}
This yield surface is based on the elastic criterion proposed by \cite{francois_08}. Its shape, smooth and convex, can be regarded as a softened approximation of the Von Mises and Rankine criterions. It is in good agreement with the elastic limit identified by \cite{kupfer_69} in biaxial testings (Fig.~\ref{SSFrancois}). The first member corresponds to the Von Mises expression; the second one, due to a tensor exponential, growths quickly for positive values of any principal stress. The material constants are $\sy$, that principally rules the limit stress in compression and $\so$, that strongly influences the tension to compression stress ratio. The functions $g(d)$ and $h(d)$, with $g(0)=h(0)=1$, are hardening functions (whose detailed expressions are given by Eq.~\ref{hardfun1} and \ref{hardfun2}).
\begin{equation}
f(\SIG,d) = \frac{||\SIGD||}{g(d)} + \frac{\so}{h(d)} \left( \left\| \exp \left( \frac{\SIG}{\so} \right) \right\| - \sqrt{3} \right) - \sy.
\label{yieldsurf}
\end{equation}
Naming $(\si, \sii, \siii)$ the eigenvalues of the stress tensor and considering sound material $(d=0)$, this criterion writes:
\begin{eqnarray}
\tilde{f}(\si,\sii,\siii,d=0) &=& \sqrt{ \frac{(\sii-\siii)^2+(\siii-\si)^2+(\si-\sii)^2} {3} }\nonumber\\
&+& \so \left( \sqrt{ \exp \left(\frac{2 \si}{\so}\right) + \exp \left(\frac{2 \sii}{\so}\right) + \exp \left(\frac{2 \siii}{\so}\right) } - \sqrt{3} \right) - \sy.
\label{eigenyieldsurf}
\end{eqnarray}
The identification of $(\so,\sy)$ is done with respect to the measured elastic limit stresses $\sco=-9.98$ MPa in pure compression and $\ste=1.29$ MPa in pure tension. The difference $\tilde{f}(\ste,0,0,0)-\tilde{f}(\sco,0,0,0)$ gives:
\begin{equation}
\so \sqrt{\exp \left( \frac{2 \ste}{\so} \right) + 2}
+ \ste \sqrt{\frac{2}{3}} =
\so \sqrt{\exp \left( \frac{2 \sco}{\so} \right) + 2}
- \sco \sqrt{\frac{2}{3}}.
\end{equation}
This equation accepts a numeric resolution given by the intersection of the two functions of $\so$ that correspond to left and right members. This intersection is unique because each one is monotonic. The value of $\sy$ is obtained from $\tilde{f}(\ste,0,0,0)=0$ or $\tilde{f}(\sco,0,0,0)=0$. Numerical values of $(\so,\sy)$ are reported in table (\ref{tabcons}).
%
%
%

\textbf{Tangent modulus and evolution equation.}
In case of dissipative transformation, the consistency equation $\d f(\SIG,d)=0$ and the differential of $\SIG(\EPS,d)$ give the damage evolution $\d d$:
\begin{equation}
\d d = - 
\left( \frac{\partial f}{\partial d} + \frac{\partial f}{\partial \SIG} : \frac{\partial \SIG}{\partial d} \right)^{-1}
 \frac{\partial f}{\partial \SIG} : \TENS{H}^0 :  \d \EPS\label{evold}.
\end{equation}
and the expression of the tangent modulus $\TENS{H}=\d \SIG / \d \EPS$:
\begin{equation}
\TENS{H}=\left[ \TENS{I}+\left(\frac{\partial f}{\partial d}\right)^{-1}\frac{\partial \SIG}{\partial d}\otimes\frac{\partial f}{\partial \SIG} \right]^{-1}:\TENS{H}^0\label{H}.
\end{equation}
The inverse of the fourth rank tensor can easily be obtained in a convenient tensorial base \cite{rychlewski_84} as the inverse of a 6x6 square matrix. The involved derivatives are obtained without difficulty from Eq.~\ref{sigh}, \ref{sigd} and \ref{yieldsurf}:
\begin{eqnarray}
\frac{\partial f}{\partial d}&=&-\frac{||\SIGD||g'(d)}{g^2(d)}-\frac{\so h'(d)}{h^2(d)}\left( \left\| \exp \left( \frac{\SIG}{\so} \right) \right\| - \sqrt{3} \right)\label{dfdd},\\
\frac{\partial f}{\partial \SIG}&=&\frac{\SIGD}{g(d) ||\SIGD||}+\frac{1}{h(d)}\frac{\exp(2\SIG/\so)}{|| \exp(\SIG/\so) ||}\label{dfdS},\\
\frac{\partial \SIG}{\partial d}&=&-2\mu\varphi(\ed)^2 \frac{\ID}{\sqrt{3}}-2\mu(1+2\varphi\eh)\EPSD\label{dSdd}.
\end{eqnarray}
The elasticity equations (\ref{sigh}, \ref{sigd}), the yield condition (\ref{yieldsurf}) and the evolution equations (\ref{evold} to \ref{dSdd}) constitute the set that is required for the use of the model within a FEM code.
%
%
\section{Tension and compression simulation}\label{sectenscomp}
%
%

\textbf{Stress to strain relationship.}
The stress tensor expresses as $\SIG=\sigma\vec{e}_1\otimes\vec{e}_1$. Due to the isotropy of both the material and the model, the strain tensor writes:
\begin{equation}
\EPS = \left[\begin{array}{ccc}\varepsilon & 0 & 0 \\0 & -\nut\varepsilon & 0 \\0 & 0 & -\nut\varepsilon \end{array}\right]\label{32}.
\end{equation}
where $\varepsilon$ is the uni-dimensional strain and $\nut$ the apparent Poisson's ratio. Using the projection equations (\ref{defD}, \ref{projections}) and denoting $\tens{D}=\EPSD/|| \EPSD ||$ (see Appendix \ref{tensoform}), we have:
\begin{eqnarray}
\tens{D} &=& \textrm{sign} (\varepsilon(1+\nut))
\left[\begin{array}{ccc}2/\sqrt{6} & 0 & 0 \\0 & -1/\sqrt{6} & 0 \\0 & 0 & -1/\sqrt{6}\end{array}\right]\label{dtrac},\\
\eh&=&\frac{1-2\nut}{\sqrt{3}}\varepsilon, \quad \ed=\sqrt{\frac{2}{3}} \left| \varepsilon (1+\nut) \right|\label{ehdtrac}.
\end{eqnarray}
Using these equations in Eq.~\ref{sigh1} and \ref{sigd1} leads to two expressions for the stress to strain relationship:
\begin{eqnarray}
\sigma &=& 3K(1-2\nut)\varepsilon-\frac{4\mu\varphi d}{\sqrt{3}}(1+\nut)^2\varepsilon^2\label{sigh2},\\
\sigma &=& 2 \mu \left( 1-d-2\varphi d\frac{1-2\nut}{\sqrt{3}} \varepsilon \right)(1+\nut)\varepsilon\label{sigd2}.
\end{eqnarray}
Eliminating $\sigma$ between them gives a second order equation in $\nut$:
\begin{equation}
\nut^2 \frac{12\mu\varphi d}{\sqrt{3}} \varepsilon+
\nut \left[ 2\mu(1-d)+6K+\frac{12\mu\varphi d \varepsilon}{\sqrt{3}} \right]+
2\mu(1-d)-3K=0
\label{nutilde}.
\end{equation}
Eq.~\ref{sigh2} or \ref{sigd2} gives from zero to two solutions $\sigma$ for a given strain $\varepsilon$ (these solutions correspond to different values of $\nut$, then to different strain tensors $\EPS$).%
%

\textbf{Dissipative transformations equations.}
The yield function (\ref{yieldsurf}) becomes:
\begin{equation}
f^*(\sigma,d)=\sqrt{\frac{2}{3}} \frac{| \sigma |}{g(d)} + \frac{\so}{h(d)} \left( \sqrt{\exp \left( \frac{2 \sigma}{\so} \right) + 2} -\sqrt{3} \right)-\sy.
\label{oneDyieldsurf}
\end{equation}
As in the general 3D case, from the differential of $\sigma(\varepsilon,d)$ and the consistency equation $\d f^*(\sigma,d)=0$ is obtained the stress to strain differential relationship:
\begin{equation}
\d \sigma \left( \frac{\partial f^*}{\partial d} + \frac{\partial f^*}{\partial \sigma} \frac{\partial \sigma}{\partial d} \right) = \frac{\partial f^*}{\partial d} \frac{\partial \sigma}{\partial \varepsilon} \d \varepsilon\label{de1d}.
\end{equation}
From Eq.~\ref{oneDyieldsurf} we have:
\begin{eqnarray}
\frac{\partial f^*}{\partial \sigma} &=& \sqrt{\frac{2}{3}} \frac{\textrm{sign}(\sigma)}{g(d)} + 
\frac{ \exp\left(\frac{2\sigma}{\so}\right)} {h(d) \sqrt{2+\exp\left(\frac{2\sigma}{\so}\right)}} \label{dfds1d},\\
\frac{\partial f^*}{\partial d} &=& - \sqrt{\frac{2}{3}} \frac{| \sigma | g'(d)}{g^2(d)} -
\frac{\so h'(d)}{h^2(d)} \left( \sqrt{2+\exp\left(\frac{2\sigma}{\so}\right)} - \sqrt{3} \right) \label{dfdd1d}.
\end{eqnarray}
From the stress to strain relationship (Eq.~\ref{sigh2},\ref{sigd2} and \ref{nutilde}), the following derivatives, in which $\partial \sigma / \partial \varepsilon$ is the tangent Young modulus and $\Delta$ the discriminant (in Eq.~\ref{nutilde}), are obtained:
\begin{eqnarray}
\frac{\partial \sigma}{\partial \varepsilon}  &=&
3K(1-2\nut)-\frac{8}{\sqrt{3}}\mu\varphi d (1+\nut)^2 \varepsilon - \left( 3K + \frac{4}{\sqrt{3}}\mu\varphi d (1+\nut) \varepsilon \right) 2\varepsilon \frac{\partial \nut}{\partial \varepsilon}\label{dsde1d},\\
\frac{3}{2\mu\varepsilon(1+\nut)} \frac{\partial \sigma}{\partial d}  &=& \frac{2\varphi}{\sqrt{3}} \left( 2 \varepsilon \frac{\partial \nut}{\partial \varepsilon} - 1 + 2 \nut \right)(1+\nut)\varepsilon- \left( 1+2\varphi \frac{1-2\nut}{\sqrt{3}}\varepsilon \right) \left(2(1+\nut)+2\varepsilon\frac{\partial \nut}{\partial \varepsilon}\right)\label{dsdd1d},\\ 
2 \varepsilon \frac{\partial \nut}{\partial \varepsilon} &=& -(1+2\nut) \pm \frac{-2\mu(1-d)+12\mu\varphi d \varepsilon / \sqrt{3}}{\sqrt{\Delta}}\label{dnude1d}.
\end{eqnarray}
%
%

\textbf{Localization in tension and compression.}
In the tension or compression state, the axis 1 is the symmetry axis. As axis 2 and 3 have an equivalent role, we can define with no restriction $\vec{n}$ in the plane $[\vec{e}_1,\vec{e}_2]$ as: $\vec{n}=\cos(\theta)\vec{e}_1+\sin(\theta)\vec{e}_2$.
For dissipative transformations, the general equations of Sec.~\ref{dissbe} allow to compute $\det(\vec{n}.\TENS{H}.\vec{n})$ numerically by searching its minimum value with respect to $\theta$.
The non convex potential authorizes possible localization during elastic transformations. From Eq.~\ref{H0} we have:
%
%
\begin{equation}
\vec{n}.\TENS{H}^0.\vec{n} =
\left[\begin{array}{ccc}
b+(a+4c)\cos^2(\theta) & (a+c)\sin(\theta)\cos(\theta) & 0 \\
(a+c)\sin(\theta)\cos(\theta) & b+(a-2c)\sin^2(\theta) & 0 \\
0 & 0 & b
\end{array}\right]\nonumber,
\end{equation}
\begin{equation}
a=K+\frac{\mu}{3}(1-d)-\frac{2}{3 \sqrt{3}}\mu\varphi d (1-2\nut) \varepsilon, \quad
b=\mu(1-d)-\frac{2}{\sqrt{3}}\mu\varphi d (1-2\nut) \varepsilon, \quad
c=- \frac{4}{3\sqrt{3}}\mu\varphi d (1+\nut) \varepsilon
\label{nhonund}.
\end{equation}

%

\begin{figure}[htbp]
	\begin{center}
		\includegraphics[scale=0.7]{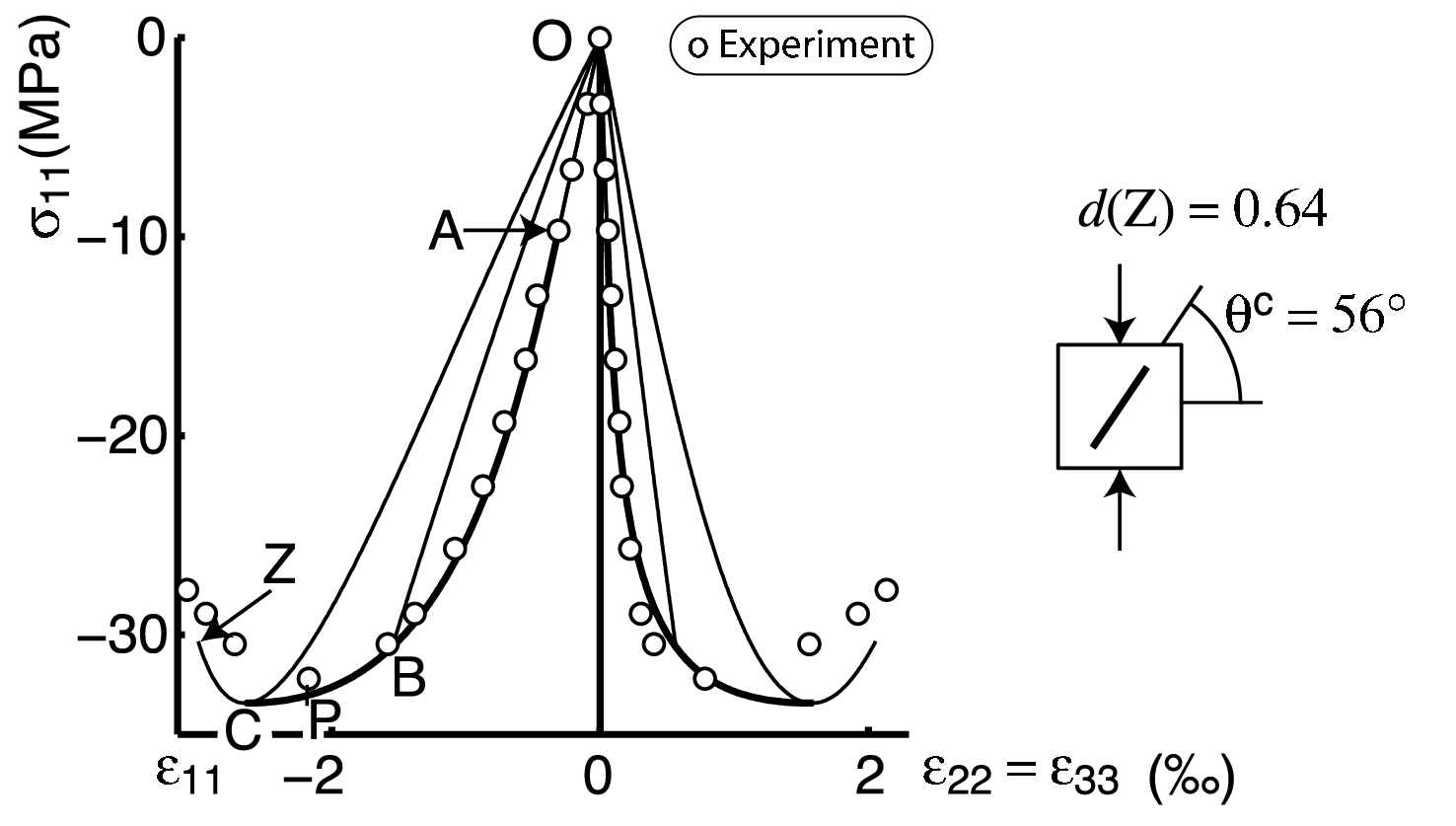}
		\caption{Uniaxial compression simulation}
		\label{compression}
	\end{center}
\end{figure}

\textbf{Identification.}
The identification of the Young Modulus $E=31.2$ GPa and the Poisson's ratio $\nu=0.162$ being very classic, they will not be detailed; the corresponding bulk and shear moduli $(K, \mu)$ are given in table (\ref{tabcons}). The procedure of identification of $(\so,\sy)$ is given in Sec.~\ref{dissbe}. Eliminating $d$ between Eq.~\ref{sigh2} and \ref{sigd2} gives the following expression that can be used with the experimental data in order to determine $\varphi$:
\begin{equation}
\varphi = \frac{\sqrt{3}}{2}
\frac{\sigma / \varepsilon-3K(1-2\nut)}
{3K\varepsilon(1-2\nut)^2 - 2\mu\varepsilon(1+\nut)^2 + \sigma\nut}.
\end{equation}
The hardening functions $g(d)$ and $h(d)$ are identified with respect to the transformation AC (see Fig.~\ref{compression}). Their structure has been chosen in order to have a smooth transition at the yielding point A.
\begin{eqnarray}
g(d) &=& 1+g_0 \sqrt{d},\label{hardfun1}\\
h(d) &=& 1-d+h_0 (\sqrt{d}-d)\label{hardfun2}.
\end{eqnarray}
\begin{table}[htdp]
\caption{Material constants}
\begin{center}
\begin{tabular}{|cc|ccc|cc|}
\hline
$K$ & $\mu$ & $\sy$ & $\so$ & $\varphi$ & $g_0$ & $h_0$ \\
\hline
15.4 & 13.4 & 8.004 & 0.4551 & 160 & 3.00 & 160 \\
(GPa) & (GPa) & (MPa) & (MPa) & & & \\
\hline
\end{tabular}
\end{center}
\label{tabcons}
\end{table}%
%
%

\begin{figure}[htbp]
	\begin{center}
	\includegraphics[scale=0.7]{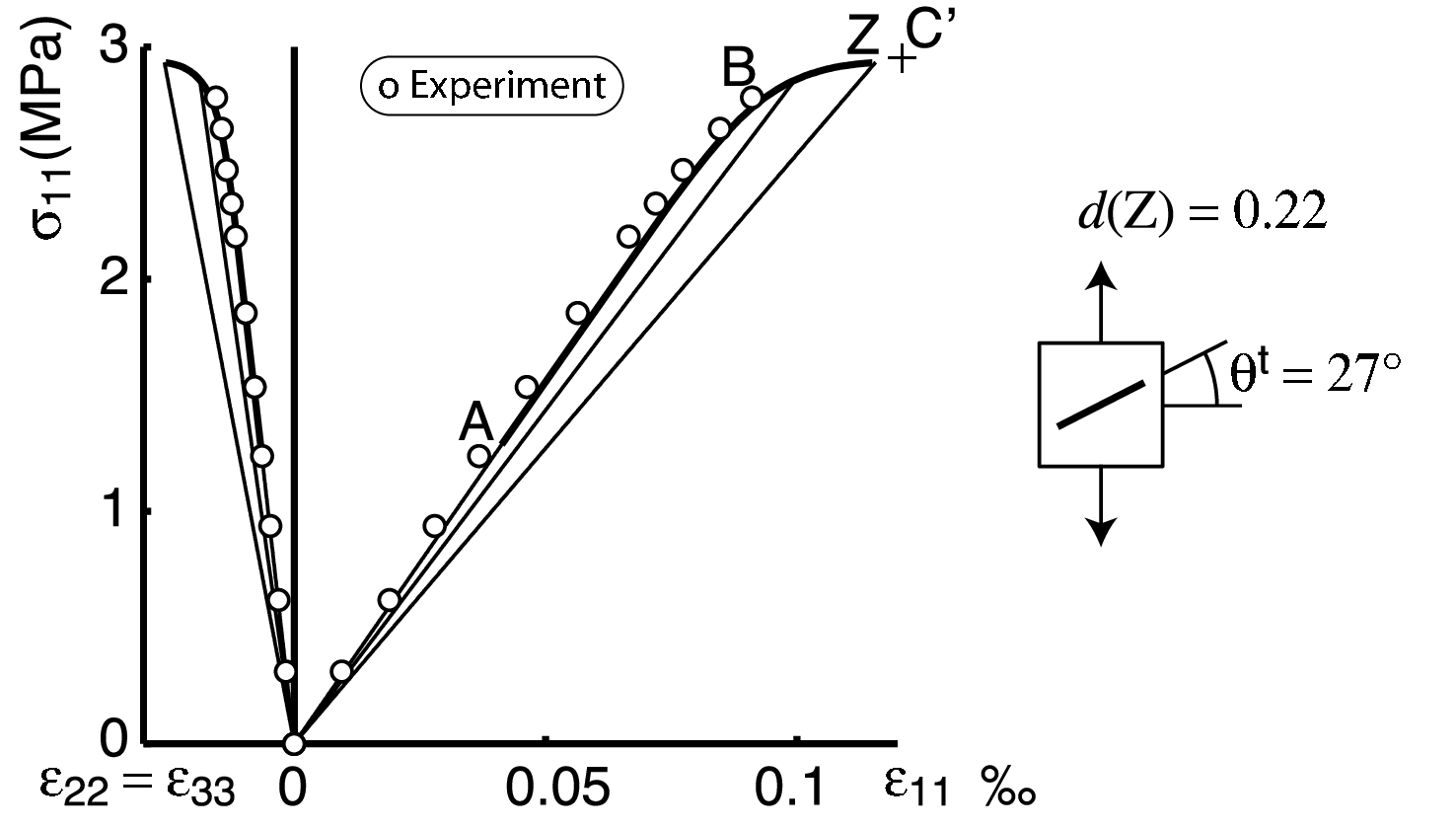}
	\caption{Uniaxial tension simulation}
	\label{tension}
	\end{center}
\end{figure}

\textbf{Uniaxial compression simulation.}
The set of equations (\ref{32} to \ref{dnude1d}) defines the strain driven compression and tension curves (Fig.~\ref{compression} and \ref{tension}), allowing the comparison with Kupfer's data (circles).
%
{The compression starts by the linear elastic transformation OA, with $d=0$.
The material begins to yield at point A where $\sigma$(A)= $\sco$.}
{The transformation ABC corresponds to the damage growth, up to point C where $d$(C)$= 0.64$. The localization criterion is never reached during ABC. At point P, $\nut=0.5$: the material begins to dilate.}
{Unloadings such as OBO for $d=0.5$ and OCO for $d=0.64$ reveal weak  nonlinear elasticity: it can be noticed that the curvature of OC (the tangent modulus decreases as the load increases) is qualitatively in good agreement with the experiments of \cite{ramtani_90} for example. Of course hysteresis loops or permanent strains are not described by the present simple form of the model that does not takes into account plasticity effects.}
{Point C corresponds to a horizontal slope for the transformation. Then $\d \sigma=0$ and, as the value of $\partial f / \partial d \neq 0$ at this point, Eq.~\ref{de1d} shows that $\partial \sigma / \partial \varepsilon=0$: the elastic transformation OCZ has a also an horizontal tangent at point C.}
{After C, $\partial \sigma / \partial \varepsilon<0$, the transformation CZ is a non linear elastic softening. The damage remains fixed at $d(C)$. From some point just before Z, an unloading follows the unrealistic elastic transformation ZCO (this aspect of the model remains to be enhanced).}
{At point Z the localization condition relative to the elastic transformations (Eq.~\ref{nhonund}), is reached and this point is close to the experimental breakdown point identified by Kupfer (although this value has not been used during the identification process).}
%
{The continuation of the elastic transformation after Z is, for information, drawn on Fig.~\ref{planDP}; as $\eh$ and $\ed$ are dependent upon $\varepsilon$ by Eq.~\ref{ehdtrac}, they become simultaneously equal to zero at the saddle point H.}
The localization arises at point Z for the angle $\theta^c=56$ degrees: the crack plane is rather close to the load axis $\vec{e}_1$, as classically observed in experiments.
The apparent Poisson's ratio $\nut$ reaches 0.70 at the point Z and this large value, that cannot be reached by a classic damage model, is responsible of the good description of the transverse strain.
%

\textbf{Uniaxial tension simulation.}
The tension curve (Fig.~\ref{tension}) is also referred to points A,B,C',Z whose roles are similar to the ones used in compression. 
%
%
{The yielding occurs at point A, at the identified value $\ste$ that is, according to Kupfer and confirmed by \cite{terrien_80}, less than the half of the peak stress: this result is different from the classic vision of a fragile behavior in tension.}
{At point C', $\partial f / \partial d = 0$ then (from Eq.~\ref{de1d}) the slope is horizontal. It is not reached because the localization occurs before, at point Z, during the dissipative transformation.}
%
If \cite{terrien_80} describes a post-peak evolution, Kupfer does not, neither this model with the retained constants. 
Again, although not used in the identification process, the stress $\sigma$(Z) is close to the experimental value.
The angle of localization at point Z is $\theta^t=27$ degrees, leading to a crack plane rather close to be orthogonal to the tensile axis $\vec{e}_1$, as commonly observed (moreover most of damage models suppose a crack at 0 degree).
The lateral strain, due to low damage values, remains very small, consistently with experiments.
The damage value at the localization point $d$(Z)$=0.22$ is weak: although isotropic, this model allows to have a rupture in tension and to keep a loading capacity in compression.
%
%
\section{Equi-biaxial simulations}\label{secbi}
In this section we consider an equi-biaxial loading (strain driven) in order to validate the model with the constants identified in tension and compression. The stress tensor is $\SIG=\sigma(\vec{e}_2\otimes\vec{e}_2+\vec{e}_3\otimes\vec{e}_3)$. Due to the initial isotropy of both the material and the model, the strain tensor writes:
\begin{equation}
\EPS = \left[\begin{array}{c c c}-\nust\varepsilon & 0 & 0 \\0 & \varepsilon & 0 \\0 & 0 & \varepsilon \end{array}\right].
\end{equation}
with $\varepsilon\leqslant0$. The projection equations (\ref{defD}, \ref{projections}) give:
%
\begin{eqnarray}
\tens{D} &=& \textrm{sign} (\varepsilon(1+\nust))
\left[\begin{array}{ccc}-2/\sqrt{6} & 0 & 0 \\0 & 1/\sqrt{6} & 0 \\0 & 0 & 1/\sqrt{6}\end{array}\right]\label{dbi},\\
\eh&=&\frac{2-\nust}{\sqrt{3}}\varepsilon, \quad \ed=\sqrt{\frac{2}{3}} \left| \varepsilon (1+\nust) \right|\label{ehdbi}.
\end{eqnarray}
Using these equations in Eq.~\ref{sigh1} and \ref{sigd1} gives two expressions for the stress to strain relationship:
\begin{eqnarray}
\sigma &=& \frac{3K}{2}(2-\nust)\varepsilon - \frac{2 \mu \varphi d}{\sqrt{3}} (1+\nust)^2\varepsilon^2,\\
\sigma &=& 2 \mu \left(1-d-2\varphi d \frac{2-\nust}{\sqrt{3}} \varepsilon \right) (1+\nust) \varepsilon.
\end{eqnarray}
Eliminating $\sigma$ gives the following second order equation which accepts from zero to two solutions in $\nust$:
\begin{equation}
\nust^2 2\sqrt{3}\mu\varphi d \varepsilon+
\nust \left( 2\mu(1-d)+\frac{3K}{2} \right)
+2\mu(1-d)-3K- 2\sqrt{3}\mu\varphi d \varepsilon=0.
\end{equation}
Other calculi are similar to the previous case.
%

\textbf{Equi-biaxial compression simulation.}
\begin{figure}[hbtp]
	\begin{center}
	\includegraphics[scale=0.7]{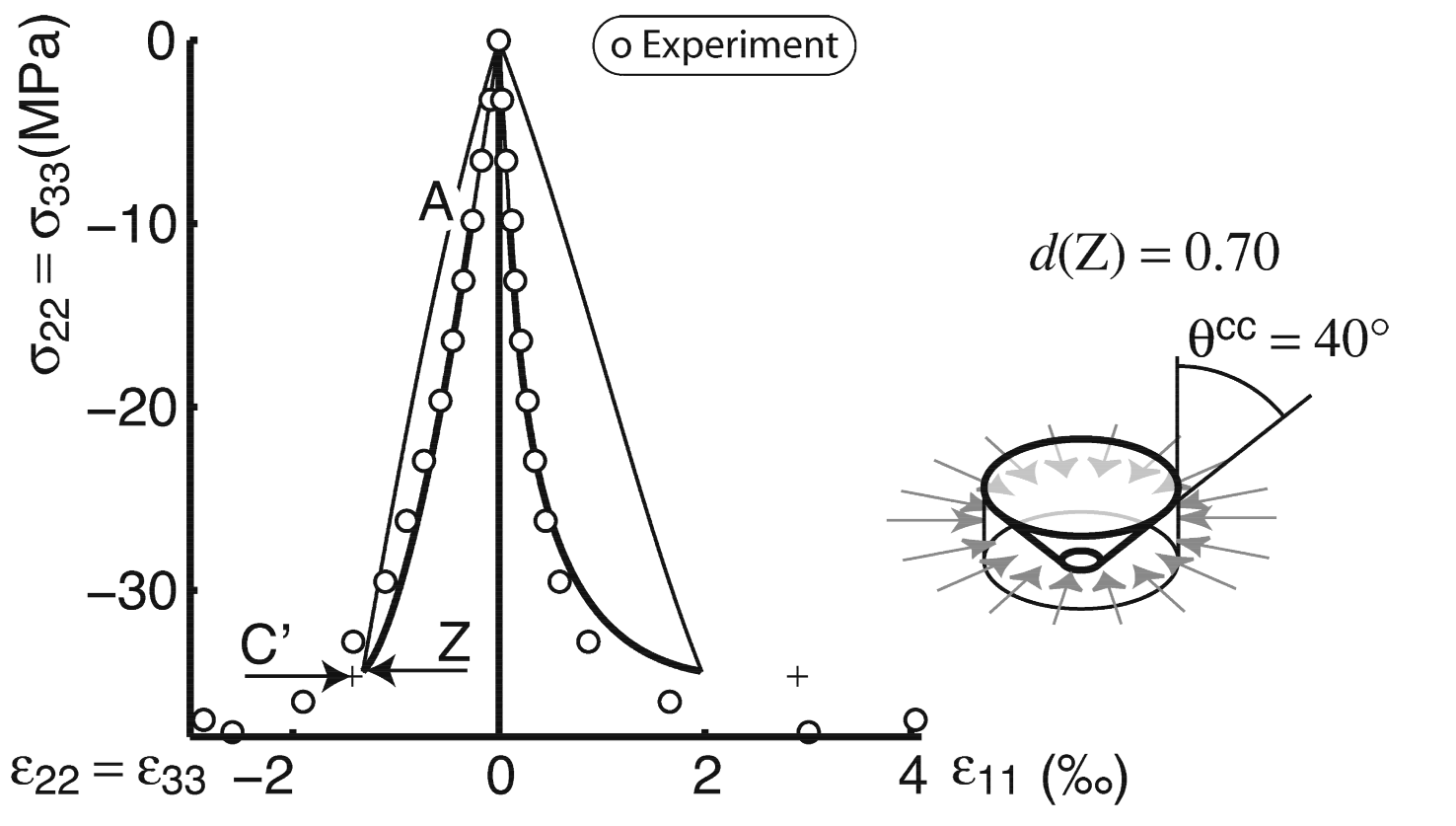}
	\caption{Equi-biaxial compression simulation}
	\label{biaxialc}
	\end{center}
\end{figure}
{Directly dependent upon the yield surface, the yield point A (Fig.~\ref{biaxialc}) is well depicted.}
{The damage increases during transformation AZ up to $d$(Z)$=0.70$, this value being greater than the maximal one obtained in uniaxial compression.}
{At point Z, (which is before the point C' where $\partial f / \partial d = 0$) the localization condition for dissipative transformations (Eq.~\ref{detnhn}) is reached.}
%
The localization angle is $\theta^{cc}=40$ degrees, consistently with Kupfer's \emph{post mortem} picture. The conical crack shape in Fig.~\ref{biaxialc} has been chosen in order to respect the symmetry axis of the problem. Without contradiction, the observed structure is pyramidal, this shape allowing the kinematics of the concrete pieces.
The localization appears too quickly and the large dilatancy is a little overestimated. 
Moreover, this post peak region differs according to different experimental methods and concretes: for example the high strength concrete tested by \cite{hussein_00} exhibits no post peak.
%

\textbf{Equi-biaxial tension simulation.}
\begin{figure}[h!]
	\begin{center}
	\includegraphics[scale=0.7]{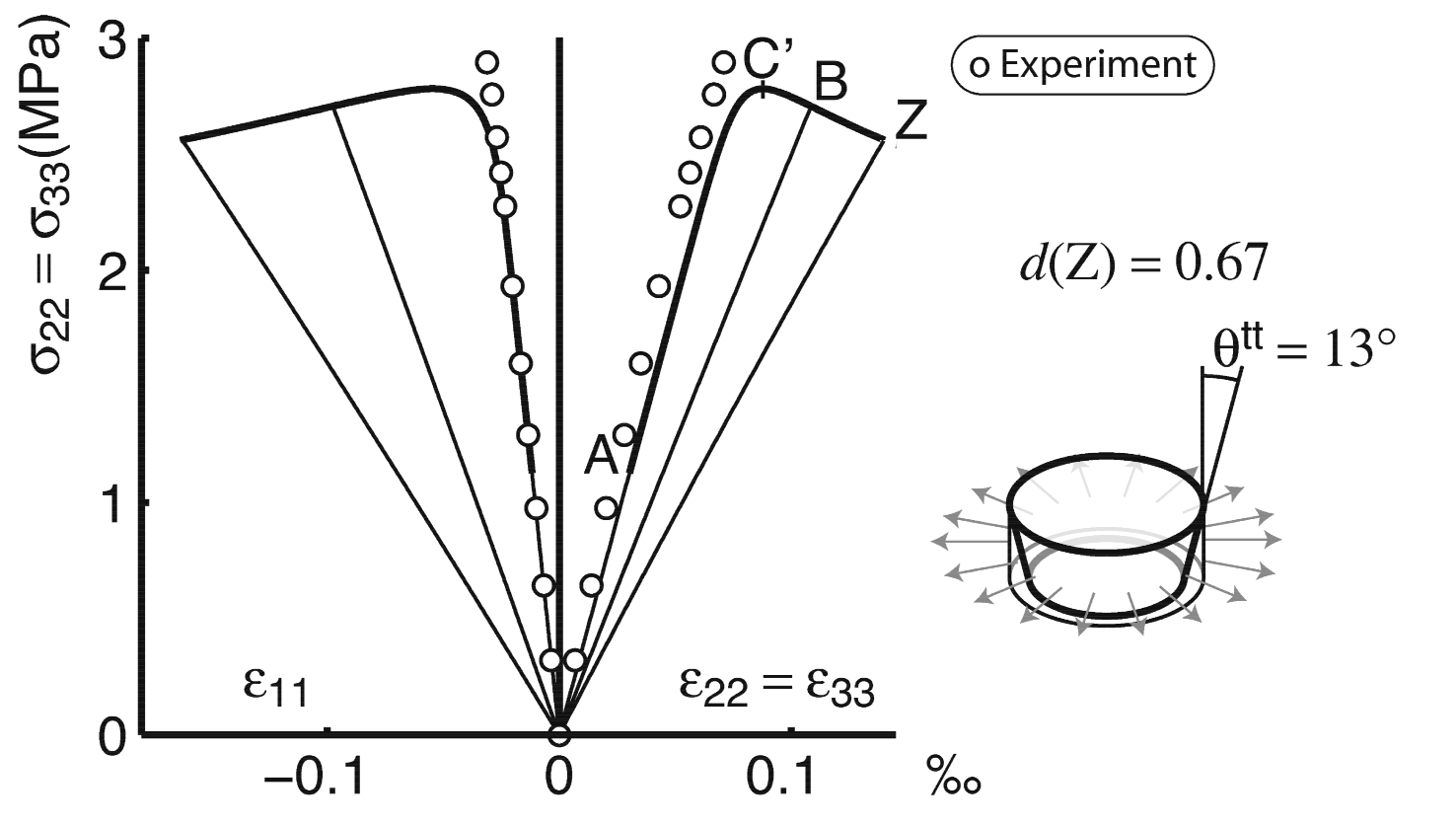}
	\caption{Equi-biaxial tension simulation}
	\label{biaxialt}
	\end{center}
\end{figure}
%
{The part before the peak stress is in good agreement with the experiments for both axial and lateral strains (Fig.~\ref{biaxialt}).}
{Contrary to experiments, a dissipative post-peak evolution C'Z is described (it represents a minor problem compared to the elastic transformation CZ in the uniaxial compression case).}
The damage $d$(Z)$=0.67$ at ultimate point reaches about the same value than in biaxial tension.
The localization angle $\theta^{tt}=12.6$ degrees obtained in this case involves cracks approximatively orthogonal to the load directions, very consistent with Kupfer's \emph{post mortem} pictures. Again, the conical shape is only the representation of the symmetry of the problem as the model only defines the angle of the crack, not its shape.%

\newpage

\section{Conclusions}


This study shows that the simple coupling introduced between the hydrostatic and deviatoric parts of the strain in the free energy leads to a number of consequences, even if the primary effect, the non linear elasticity, slightly influences the shape of the elastic discharges. 
The important dilatancy of the concrete at ultimate stages is well depicted. 
The localization criterion is used as an indicator of the creation of a macroscopic crack, it applies at stress and strain levels that are quite consistent with experiments.
The damage value predicted at localization in pure tension remains small, letting to the material some load bearing capacity, for example for further compression.
The crack angle given by the localization condition is different in tension and compression and consistent with observations.

The description of these effects constitutes an improvement with respect to most of available damage models of comparable complexity (five constants are used). They are particularly relevant when, as in earthquake engineering, the response of a concrete structure under severe conditions is considered. 
The global structure of the model and its embedding thermodynamic framework, opens the way to numerical implementation within a finite element code.


The introduction of a plasticity formalism \cite{ragueneau_00,salari_04} would help to describe missing effects such as hysteresis loops and permanent strains and to correct some irrelevant post-peak responses. The hydrostatic part of the free energy could be affected by damage in order to extent the field of application of the model to confined states and avoid the damage locking condition present in the actual form.
The three-dimensionnal stress to strain relationship is univocal, however the inverse one is not: interesting theoretical analysis can be done on stress driven loadings as they may involve multi phased states (this refers to the work of \cite{eriksen_75} in elasticity and \cite{froli_00,puglisi_05} in plasticity).


%

\section*{Appendices}

\appendix

%
%
\section{Tensorial formalism}\label{tensoform}
For any symmetric second order tensor $\tens{A}$ the hydrostatic and deviatoric parts are denoted respectively by $\tens{A}\e{h}$ and $\tens{A}\e{d}$. The hydrostatic part is obtained by the projection onto the normed hydrostatic tensor $\ID/\sqrt{3}$ (where $\ID$ is the identity tensor).
\begin{eqnarray}
\tens{A} &=& \tens{A}\e{h} + \tens{A}\e{d}\label{ad}\\
\tens{A}\e{h} &=& \left( \tens{A}:\frac{\ID}{\sqrt{3}} \right) \frac{\ID}{\sqrt{3}}\label{ah}
\end{eqnarray}
The symbol "$:$" denotes the double contraction \emph{i.e.} $\tens{A}:\ID=A\i{ij}\delta\i{ji}=\trace(\tens{A})$. In this model, the deviatoric parts $\SIGD$ and $\EPSD$ (of the Cauchy stress $\SIG$ and the infinitesimal strain $\EPS$) will remain collinear but may be \emph{a priori} of opposite directions. We choose to define this direction with respect to the normed tensor $\tens{D}$ given by:
\begin{equation}
\tens{D} = \frac{\EPSD}{|| \EPSD ||}
\label{defD}
\end{equation}
The norm used is the euclidean natural one \emph{i.e.} $||\tens{A}||= \sqrt{A \i {ij} A \i {ij}}$. 
The algebraic values $(\eh, \ed, \sh, \sd)$ are defined as:
\begin{equation}
\eh = \EPS : \frac{\ID}{\sqrt{3}},\quad
\sh = \SIG : \frac{\ID}{\sqrt{3}},\quad
\ed = \EPS : \tens{D},\quad
\sd = \SIG : \tens{D}
\label{projections}
\end{equation}
One can remark that $\ed$ is positive but $\sd$ can \emph{a priori} either be positive or negative. These projections are known under the names $p$ and $q$ where $p=-\sh/\sqrt{3}$ and $q=\sqrt{3/2}|\sd|$ in soils mechanics.

\section{Useful property of the deviator} \label{calcbound}
The sorted eigenvalues of $\EPSD$ are $(\ed\i{I},\ed\i{II},\ed\i{III})$. From Eq.~\ref{projections}, we have:
\begin{eqnarray}
	(\ed\i{I})^2+(\ed\i{II})^2+(\ed\i{III})^2&=&(\ed)^2\\
	\ed\i{I}+\ed\i{II}+\ed\i{III}&=&0\\
	\ed\i{I}\geqslant\ed\i{II}\geqslant\ed\i{III}&&
\end{eqnarray}
This leads to the following expressions:
\begin{eqnarray}
	\ed\i{I} = -\frac{\ed\i{II}}{2}+\frac{1}{2}\sqrt{2(\ed)^2-3(\ed\i{II})^2} &,&
	\ed\i{III} = -\frac{\ed\i{II}}{2}-\frac{1}{2}\sqrt{2(\ed)^2-3(\ed\i{II})^2}
\end{eqnarray}
As soon as $\ed\i{II}$ evolves between its bounds $[-\ed/\sqrt{6};\ed/\sqrt{6}]$ (given respectively by $\ed\i{II}=\ed\i{III}$ and $\ed\i{II}=\ed\i{I}$), these functions are monotonic and:
\begin{equation}
	\sqrt{\frac{2}{3}}\ed\geqslant\ed\i{I}\geqslant\frac{\ed}{\sqrt{6}}\label{epsbound}
\end{equation}
%


\bibliographystyle{unsrt}
\bibliography{mybibli}

\begin{thebibliography}{10}

\bibitem{ouglova_06}
Anna Ouglova, Yves Berthaud, Marc Fran{\c c}ois, and Fran{\c c}ois Foct.
\newblock Mechanical properties of ferric oxide formed by corrosion in
  reinforced concrete structures.
\newblock {\em Corrosion Sci.}, 48:3988--4000, 2006.

\bibitem{jamet_84}
P~Jamet, Alain Millard, and G~Nahas.
\newblock Triaxial behaviour of a micro-concrete complete stress-strain for
  confining pressures ranging from 0 to 100 {MPa}.
\newblock In {\em Proc. of Int. Conf. on Concrete under Multiaxial Conditions},
  volume~1, pages 133--140, Toulouse, France, 1984. Universit{\'e} Paul
  Sabatier.

\bibitem{kupfer_69}
Helmut Kupfer, Hubert~K Hilsdorf, and Hubert Rusch.
\newblock Behavior of concrete under biaxial stresses.
\newblock {\em ACI Journal}, 66:656--666, 1969.

\bibitem{rice_75}
J~W Rudnicki and J~Rice.
\newblock Conditions for the localization of deformation in pressure-sensitive
  dilatant materials.
\newblock {\em J. Mech. Phys. Solids}, 23:371--394, 1975.

\bibitem{comi_95}
Claudia Comi, Yves Berthaud, and Ren{\'e} Billadon.
\newblock On localization in ductile-brittle materials under compressive
  loadings.
\newblock {\em Eur. J. Mech. A/Solids}, 14:19--43, 1995.

\bibitem{le_05}
Tuan~Hung Le.
\newblock {\em Contribution {\`a} la mod{\'e}lisation du comportement des
  b{\'e}tons}.
\newblock Phd thesis, Universit{\'e} Pierre et Marie Curie Paris VI, 2005.

\bibitem{rychlewski_84}
J~Rychlewski.
\newblock On hooke's law.
\newblock {\em Prikl. Matem. Mekhan.}, 48:420--435, 1984.

\bibitem{ladeveze_93}
Pierre Ladeveze.
\newblock On anisotropic damage theory.
\newblock {\em Failure criteria of Structured media, Proc. of the CNRS
  international colloquium No 351}, pages 355--363, 1993.

\bibitem{francois_08}
Marc Fran{\c c}ois.
\newblock A new yield criterion for the concrete materials.
\newblock {\em C. R. M{\'e}canique}, DOI: 10.1016/j.crme.2008.01.010:under
  press, 2008.

\bibitem{lemaitre_96}
Jean Lemaitre.
\newblock {\em A Course on Damage Mechanics}.
\newblock Springer, Heidelberg, 1996.

\bibitem{halphen_75}
Bernard Halphen and Quoc~Son Nguyen.
\newblock Sur les mat{\'e}riaux standards g{\'e}n{\'e}ralis{\'e}s.
\newblock {\em Journal de M{\'e}canique}, 14:39--63, 1975.

\bibitem{badel_07}
Pierre~Bernard Badel, Vincent Godard, and Jean-Baptiste Leblond.
\newblock Application of some anisotropic damage model to the prediction of the
  failure of some complex industrial concrete structure.
\newblock {\em International Journal of Solids and Structures}, 44:5848--5874,
  2007.

\bibitem{yazdani_88}
S~Yazdani and H~Schreyer.
\newblock An anisotropic damage model with dilatation for concrete.
\newblock {\em Mech. Materials}, 7:231--244, 1988.

\bibitem{francois_05}
Marc Fran{\c c}ois and Gianni {Royer Carfagni}.
\newblock Structured deformation of damaged continua with cohesive-frictional
  sliding rough fractures.
\newblock {\em Eur. J. Mech. A/Solids}, 24:644--660, 2005.

\bibitem{burlion_97}
Nicolas Burlion.
\newblock {\em Compaction des b{\'e}tons : {\'e}l{\'e}ments de mod{\'e}lisation
  et caract{\'e}risation exp{\'e}rimentale}.
\newblock Phd thesis, ENS de Cachan, 1997.

\bibitem{ramtani_90}
Salah Ramtani.
\newblock {\em Contribution {\`a} la mod{\'e}lisation du comportment multiaxial
  du b{\'e}ton endommag{\'e} avec description du caract{\`e}re unilat{\'e}ral}.
\newblock Phd thesis, Universit{\'e} Pierre et Marie Curie Paris VI, 1990.

\bibitem{terrien_80}
Michel Terrien.
\newblock Acoustic emission and post-critical mechanical behaviour of a
  concrete under tensile stress.
\newblock {\em Bulletin de Liaison du Laboratoire des Ponts et Chaussees},
  105:65--72, 1980.

\bibitem{hussein_00}
A.~Hussein and B.~Marzouk.
\newblock Behavior of high-strength concrete under biaxial stresses.
\newblock {\em ACI Mat. J.}, 1:27--36, 2000.

\bibitem{ragueneau_00}
Fr{\'e}d{\'e}ric Ragueneau, Christian {La Borderie}, and Jacky Mazars.
\newblock Damage model for concrete like materials coupling cracking and
  friction, contribution towards structural damping: First uniaxial
  application.
\newblock {\em J. Cohesive Frictional Mat.}, 5:607--625, 2000.

\bibitem{salari_04}
M~R Salari, S~Saeb, K~J Williams, S~J Patchet, and R~C Carrasco.
\newblock A coupled elastoplastic damage model for geomaterials.
\newblock {\em Comp. Meth. Appl. Mech. Eng.}, 193:2625--2643, 2004.

\bibitem{eriksen_75}
J~L Eriksen.
\newblock Equilibrium of bars.
\newblock {\em J. Elasticity}, 5:191--202, 1975.

\bibitem{froli_00}
Maurizio Froli and Gianni {Royer Carfagni}.
\newblock A mechanical model for the elastic-pastic behavior of metallic bars.
\newblock {\em Int. J. Solids Struct.}, 37:3901--3918, 2000.

\bibitem{puglisi_05}
Giovanni Puglisi and Lev Truskinovsky.
\newblock Thermodynamics of rate-independant plasticity.
\newblock {\em J. Mech. Phys. Solids}, 53:655--679, 2005.

\end{thebibliography}


\end{document}